\documentstyle[twocolumn,aps]{revtex}

\begin{document}
\title{Continuous Unitary Transformations }
\author{Vladimir. L. Safonov \thanks{e-mail: vsafonov@ucsd.edu}}
\address{Center for Magnetic Recording Research, University of California - San\\
Diego, 9500 Gilman Drive, La Jolla, CA 92093-0401, U.S.A.}
\maketitle

\begin{abstract}
In the present time we observe a growing number of publications where the,
so-called, flow equations are successfully used to diagonalize Hamiltonians
by means of an appropriate unitary transformation. Here we discuss and
compare the flow equations (FE) method (proposed in 1994) with the method of
one step continuous unitary transformations (OS CUT) (proposed in 1982). It
is shown that the FE method can be considered as a generalization of the OS
CUT approach to the case of parameter dependent generator. The OS CUT method
gives {\it linear} differential equations for the diagonalization procedure.
In the FE method the system of differential equations is {\it nonlinear}.
Finally we discuss the generalization of idea of continuous unitary
transformations for the case of quantum equations of motion (Heisenberg
picture and density matrix).
\end{abstract}

\section{Introduction}

Unitary transformations play an extremely important role in physics. For
example, with the help of unitary transformation sometimes it is possible to
simplify a problem of interacting quasiparticles by eliminating
``inconvenient'' interaction terms from initial Hamiltonian and construct
corresponding effective interaction terms.

In the present time we observe a growing number of publications (see, \cite
{wegner0},\cite{keh},\cite{wegner1},\cite{Stein0},\cite{moca},\cite{Stein}, 
\cite{Guba},\cite{keh1},\cite{knetter},\cite{Domanski},\cite{raas},\cite{Jap}%
, \cite{white},\cite{Ran}), where the, so-called, flow equations (FE) method
is successfully used to diagonalize Hamiltonians by means of an appropriate
unitary transformation. This method of continuous unitary transformations
looks very convenient and universal. The transformed Hamiltonian appears as
a result of solving differential equations on some formal parameter.

It should be noted that quite the same idea and an analogous method
(Safonov, 1982 \cite{Safonov0}, 1983 \cite{Safonov1}) to construct nonlinear
unitary transformation was published far before the original paper on FE
method (Wegner, 1994 \cite{wegner0}) and just did not attract enough
attention of researches. For convenience, we shall call the approach of
Refs. \cite{Safonov0},\cite{Safonov1} as a one-step continuous unitary
transformation (OS CUT) method.

The goal of the present paper is to discuss and compare FE and OS CUT
methods. As it will be shown later, the FE method is a generalization of OS
CUT approach. The only difference between the methods is that the unitary
transformation generator is assumed to be dependent on a formal parameter in
the FE method and the generator does not depend on the parameter in the OS
CUT method. In the CUT method we obtain {\it linear} differential equations.
In the FE method we should solve a {\it nonlinear} system of differential
equations. Below we shall demonstrate both methods by several examples.

It is interesting that the OS CUT method idea can be applied to the case of
quantum equations of motion. Recently Mi\v{s}ta and Filip \cite{mista}\ have
developed a method of non-perturbative solutions of nonlinear Heisenberg
equations. We shall briefly discuss this approach at the end of this paper.

\section{Continuous Unitary Transformation method}

\subsection{One step continuous unitary transformation}

Here we shall formulate the key idea of the OS CUT method \cite{Safonov0}, 
\cite{Safonov1}. A general form of unitary transformation can be written as

\begin{equation}
H(\theta )=e^{\theta R}He^{-\theta R}  \label{generalUT}
\end{equation}
where$\,\theta $ is a formal parameter, $R\,$is an anti-Hermitian operator ($%
R^{\dag }=-R$), $H$ and $H(\theta )$ are the initial and transformed
Hamiltonians, respectively. The expression (\ref{generalUT}) is the solution
of the equation 
\begin{equation}
\frac{d}{d\theta }H(\theta )=[R,\,H(\theta )]  \label{equationUT}
\end{equation}
with the initial condition$\;H(0)=H$. In this form the Hamiltonian is
``rotated in an operator space around $R$.''

In order to solve this equation we should write the most general form of $%
H(\theta )$ as the expansion in terms of operator combinations with unknown $%
\theta $-dependent coefficients. The most general form of $R$ as
anti-Hermitian operator combinations also should be used.

After substituting these general (for $H(\theta )$ and $R$) expressions into
(\ref{equationUT}), one can obtain a set of {\it linear} differential
equations by comparing coefficients in analogous operator compositions in
both sides. Solving these equations with the initial conditions, we obtain
the transformed Hamiltonian $H(\theta )$. In order to eliminate
`inconvenient' terms one needs to put their coefficients (for example, for $%
\theta =1$) equal to zero. This condition defines the choice of $R$.

Usually it is simple to find the anti-Hermitian operator, which eliminates
given non-diagonal term. The OS CUT method first has been applied for the
spin Hamiltonian diagonalization \cite{Safonov0},\cite{Safonov1}. Then it
was successfully used in physics of nuclear spin waves \cite{SafNSW},
magnon-impurity interactions \cite{SafPara}, theory of superconductivity 
\cite{SafPara} and for eliminating three-boson interactions \cite{ShiSaf}.

\subsection{Flow equations for Hamiltonians}

The flow equations method \cite{wegner0} begins with a general form of a
unitary transformation:

\begin{equation}
H(\ell )=U(\ell )HU^{\dagger }(\ell ),  \label{WegUT}
\end{equation}
where $\ell $ is a continuous flow parameter. The Hamiltonian $H(\ell )$ is
transformed from the initial Hamiltonian $H(0)=H$. Differentiation (\ref
{WegUT}) yields 
\begin{equation}
\frac{d}{d\ell }H(\ell )=[\eta (\ell ),H(\ell )]  \label{FE}
\end{equation}
with the generator

\begin{equation}
\eta (\ell )=\frac{dU(\ell )}{d\ell }U^{\dagger }(\ell )=-\eta ^{\dagger
}(\ell ).  \label{WegGen}
\end{equation}
The generator $\eta (\ell )$ should be chosen in such a way, that the
off-diagonal matrix elements decay. A good choice is defined by

\begin{equation}
\eta (\ell )=[H_{d}(\ell ),H(\ell )],  \label{genUSUAL}
\end{equation}
where $H_{d}(\ell )$ is the diagonal part of $H(\ell )$ (comments \footnote{%
A rigorous criterion that this choice is most optimal seems has not been
obtained.},\footnote{%
The condition of diagonality and the condition that $\eta (\ell
)=[H_{d}(\ell ),H(\ell )]=0$ are not equivalent to each other.}).

One can see that the FE method utilizes the same key tool (first proposed in
the OS CUT method): the unitary transformation is represented as a solution
of differential equations on a formal parameter. With an accuracy of
notations the only difference between FE procedure and the OS CUT method is
the parameter dependence of the generator $\eta (\ell )$. If $\eta $ does
not depend on $\ell $ we have $U(\ell )=\exp (\eta \ell )$. This case,
obviously, exactly corresponds to the OS CUT method. If $U(\ell )=\exp
(RF(\ell ))$, where $R$ is the parameter independent anti-Hermitian operator
and $F(\ell )$ some function on $\ell $, then $\eta (\ell )=(dF(\ell )/d\ell
)R$. The Eq.(\ref{FE}) in this case can be written as

\begin{equation}
\frac{d}{(dF(\ell )/d\ell )d\ell }H(\ell )=[R,H(\ell )].  \label{bridge}
\end{equation}
This equation is reduced to Eq.(\ref{equationUT}) using a simple change of
variable $d\theta =(dF(\ell )/d\ell )d\ell $. For $U(\ell )=\exp \big(%
R_{1}F_{1}(\ell )+R_{2}F_{2}(\ell )\big)$ with commuting $R_{1}$ and $R_{2}$
anti-Hermitian operators one has $\eta (\ell )=(dF_{1}(\ell )/d\ell
)R_{1}+(dF_{2}(\ell )/d\ell )R_{2}$. The diagonalization procedure in this
case can be considered as a two transformations with $U_{1}(\ell )=\exp \big(%
R_{1}F_{1}(\ell )\big)$ and $U_{2}(\ell )=\exp \big(R_{2}F_{2}(\ell )\big)$
unitary operators. Each transformation is described by (\ref{bridge}) with $%
R_{1}$, $F_{1}(\ell )$, or $R_{2}$, $F_{2}(\ell )$.

In the most general case the $\ell $-dependent generator transformation can
not be reduced to the parameter independent case. In such a situation the FE
method can give principally different results than the OS CUT method. Let us
consider several examples.

\section{Quadratic form diagonalization}

Stein \cite{Stein0} has considered the following quadratic Hamiltonian:

\begin{equation}
H_{1}(\ell )=\sum_{{\bf q}}\Big(f_{{\bf q}}(\ell )(a_{{\bf q}}^{\dag }a_{%
{\bf q}}+b_{{\bf q}}^{\dag }b_{{\bf q}})+g_{{\bf q}}(\ell )(a_{{\bf q}%
}^{\dag }b_{-{\bf q}}^{\dag }+a_{{\bf q}}b_{-{\bf q}})\Big),
\label{HamStein}
\end{equation}
where $a_{{\bf q}}^{\dag }$, $a_{{\bf q}}$ and $b_{{\bf q}}^{\dag }$, $b_{%
{\bf q}}$ are the creation and annihilation Bose operators. Applying flow
equation

\begin{equation}
\frac{d}{d\ell }H_{1}(\ell )=[\eta _{0}(\ell ),H_{1}(\ell )]  \label{FE2}
\end{equation}
with

\begin{equation}
\eta _{0}(\ell )=\frac{1}{2}\sum_{{\bf q}}g_{{\bf q}}(\ell )(a_{{\bf q}%
}^{\dag }b_{-{\bf q}}^{\dag }-a_{{\bf q}}b_{-{\bf q}}),  \label{GenStein}
\end{equation}
the following {\it nonlinear} differential equations have been obtained:

\begin{equation}
\frac{d}{d\ell }\,f_{{\bf q}}(\ell )=-g_{{\bf q}}^{2}(\ell ),\quad \frac{d}{%
d\ell }\,g_{{\bf q}}(\ell )=-f_{{\bf q}}(\ell )g_{{\bf q}}(\ell ).
\label{EqStein}
\end{equation}
The solutions of (\ref{EqStein}) are

\begin{equation}
f_{{\bf q}}(\ell )=\epsilon _{{\bf q}}\coth (\ell \epsilon _{{\bf q}}+\ell
_{0}({\bf q})),\quad g_{{\bf q}}(\ell )=\frac{\epsilon _{{\bf q}}{\rm sgn}%
(\gamma _{{\bf q}})}{\sinh (\ell \epsilon _{{\bf q}}+\ell _{0}({\bf q}))},
\label{SteinSol}
\end{equation}
where $\epsilon _{{\bf q}}=(1-\gamma _{{\bf q}}^{2})^{1/2}$, $\ell _{0}({\bf %
q})=\frac{1}{2}\ln \left( \frac{1+\epsilon _{{\bf q}}}{1-\epsilon _{{\bf q}}}%
\right) $ and

\begin{equation}
\epsilon _{{\bf q}}=\sqrt{f_{{\bf q}}^{2}(\ell )-g_{{\bf q}}^{2}(\ell )}=%
\sqrt{f_{{\bf q}}^{2}(0)-g_{{\bf q}}^{2}(0)}  \label{SteinSpect}
\end{equation}
is the spectrum of the diagonal (at $\ell \rightarrow \infty $) Hamiltonian (%
\ref{HamStein}).

Let us now consider the analogous diagonalization procedure in the framework
of the OS CUT method. For convenience, the same notations will be used.
Simple analysis shows that the only anti-Hermitian operator that can
eliminate $a_{{\bf q}}^{\dag }b_{-{\bf q}}^{\dag }+a_{{\bf q}}b_{-{\bf q}}$
in (\ref{HamStein}) is $a_{{\bf q}}^{\dag }b_{-{\bf q}}^{\dag }-a_{{\bf q}%
}b_{-{\bf q}}$. Thus, the generator (independent on $\ell $) should be taken
in the form:

\begin{equation}
R=\frac{1}{2}\sum_{{\bf q}}G_{{\bf q}}(a_{{\bf q}}^{\dag }b_{-{\bf q}}^{\dag
}-a_{{\bf q}}b_{-{\bf q}}),  \label{MyGen}
\end{equation}
where $G_{{\bf q}}$ is unknown function independent on $\ell $. From $%
dH_{1}(\ell )/d\ell =[R,H_{1}(\ell )]$ it is simple to obtain {\it linear}
differential equations

\begin{equation}
\frac{d}{d\ell }\,f_{{\bf q}}(\ell )=-G_{{\bf q}}g_{{\bf q}}(\ell ),\quad 
\frac{d}{d\ell }\,g_{{\bf q}}(\ell )=-G_{{\bf q}}f_{{\bf q}}(\ell ).
\label{MyFG}
\end{equation}
The solution of (\ref{MyFG}) is simple:

\begin{eqnarray}
f_{{\bf q}}(\ell ) &=&\frac{f_{{\bf q}}(0)+g_{{\bf q}}(0)}{2}e^{-G_{{\bf q}%
}\ell }+\frac{f_{{\bf q}}(0)-g_{{\bf q}}(0)}{2}e^{G_{{\bf q}}\ell },
\label{Sol2} \\
g_{{\bf q}}(\ell ) &=&\frac{f_{{\bf q}}(0)+g_{{\bf q}}(0)}{2}e^{-G_{{\bf q}%
}\ell }-\frac{f_{{\bf q}}(0)-g_{{\bf q}}(0)}{2}e^{G_{{\bf q}}\ell }. 
\nonumber
\end{eqnarray}
Let the Hamiltonian (\ref{HamStein}) be diagonal at $\ell =1$: $g_{{\bf q}%
}(1)=0$. From this condition we immediately find the unknown function $G_{%
{\bf q}}$:

\begin{equation}
G_{{\bf q}}=\frac{1}{2}\ln \left( \frac{f_{{\bf q}}(0)+g_{{\bf q}}(0)}{f_{%
{\bf q}}(0)-g_{{\bf q}}(0)}\right)  \label{MyG}
\end{equation}
and the spectrum

\begin{equation}
\epsilon _{{\bf q}}=f_{{\bf q}}(1)=\sqrt{f_{{\bf q}}^{2}(0)-g_{{\bf q}%
}^{2}(0)}.  \label{MySpect}
\end{equation}
Thus, we see that in the case of exact diagonalization of quadratic
Hamiltonian the results of FE and OS CUT methods coincide with each other.

\section{Electron-phonon coupling}

Consider now the elimination of electron-phonon interaction to obtain the
effective electron-electron scattering in the framework of OS CUT and FE
methods. The Hamiltonian is (see, \cite{wegner1}):

\begin{equation}
H=H_{0}+H_{e-ph}+H_{e-e},  \label{Ex2Ham}
\end{equation}
where

\begin{equation}
H_{0}=\sum_{{\bf q}}\omega _{{\bf q}}a_{{\bf q}}^{\dag }a_{{\bf q}}+\sum_{%
{\bf k}}\varepsilon _{{\bf k}}c_{{\bf k}}^{\dag }c_{{\bf k}}
\label{freePhEl}
\end{equation}
describes free phonons ($a_{{\bf q}}^{\dag }$, $a_{{\bf q}}$) and electrons (%
$c_{{\bf k}}^{\dag }$, $c_{{\bf k}}$),

\begin{equation}
H_{e-ph}(\ell )=\sum_{{\bf k},{\bf q}}M_{{\bf k,q}}(\ell )a_{-{\bf q}}^{\dag
}c_{{\bf k+q}}^{\dagger }c_{{\bf k}}+{\rm h.c}.  \label{Ex2PhEl}
\end{equation}
describes the electron-phonon interaction and

\begin{equation}
H_{e-e}(\ell )=\sum_{{\bf k},{\bf k}^{\prime },{\bf q}}V_{{\bf k,k}^{\prime
},{\bf q}}(\ell )c_{{\bf k+q}}^{\dag }c_{{\bf k}^{\prime }-{\bf q}}^{\dagger
}c_{{\bf k}^{\prime }}c_{{\bf k}}  \label{Ex2ElEl}
\end{equation}
is the Hamiltonian of electron-electron interaction. The initial conditions
are:

\begin{equation}
M_{{\bf k,q}}(\ell =0)=M_{{\bf q}},\quad V_{{\bf k,k}^{\prime },{\bf q}%
}(\ell =0)=V_{{\bf k,k}^{\prime },{\bf q}}(0).  \label{initial}
\end{equation}

Let us first consider the diagonalization (elimination of (\ref{Ex2PhEl}))
in the framework of OS CUT method. From a simple analysis it follows that
the parameter independent generator should be taken in the form:

\begin{equation}
R=\sum_{{\bf k},{\bf q}}R_{{\bf k,q}}a_{-{\bf q}}^{\dag }c_{{\bf k+q}%
}^{\dagger }c_{{\bf k}}-{\rm h.c}.  \label{Ex2CUTgen}
\end{equation}
The diagonalization procedure gives the following {\it linear} differential
equations:

\begin{equation}
\frac{\partial M_{{\bf k,q}}(\ell )}{\partial \ell }=-\alpha _{{\bf k,q}}R_{%
{\bf k,q}}  \label{Ex2MyEq1}
\end{equation}
and for the electron pairs (with zero momentum)

\begin{equation}
\frac{\partial V_{{\bf k,-k,q}}(\ell )}{\partial \ell }=-R_{{\bf k,q}}M_{-%
{\bf k-q,q}}(\ell )-R_{-{\bf k-q,q}}M_{{\bf k,q}}(\ell ).  \label{Ex2MyEq2}
\end{equation}
From (\ref{Ex2MyEq1}) we obtain

\begin{equation}
M_{{\bf k,q}}(\ell )=M_{{\bf k,q}}(0)-\alpha _{{\bf k,q}}R_{{\bf k,q}}\ell .
\label{Msol}
\end{equation}
Let the electron-phonon Hamiltonian (\ref{Ex2PhEl}) vanishes at $\ell =1$:\ $%
M_{{\bf k,q}}(1)=0$. From this condition we immediately find

\begin{equation}
R_{{\bf k,q}}=M_{{\bf k,q}}(0)/\alpha _{{\bf k,q}}.  \label{Ex2R}
\end{equation}
Simple solution of (\ref{Ex2MyEq2}) with (\ref{Msol}), (\ref{Ex2R}) and
initial conditions (\ref{initial}) gives

\begin{equation}
V_{{\bf k,-k,q}}(1)=V_{{\bf k,-k,q}}(0)-\frac{M_{{\bf q}}^{2}\omega _{{\bf q}%
}}{\omega _{{\bf q}}^{2}-(\varepsilon _{{\bf k+q}}-\varepsilon _{{\bf k}%
})^{2}}.  \label{Vfinal}
\end{equation}
This result corresponds to the Fr\"{o}hlich's effective interaction obtained
in the second order of the perturbation theory (see, e.g., \cite{kittel}).

Let us now consider the FE method. The generator is taken in the form \cite
{wegner1}:

\begin{eqnarray}
\eta (\ell ) &=&[H_{0},H_{e-ph}(\ell )]  \label{EPHgen} \\
&=&\sum_{{\bf k},{\bf q}}M_{{\bf k,q}}(\ell )\alpha _{{\bf k,q}}a_{-{\bf q}%
}^{\dag }c_{{\bf k+q}}^{\dagger }c_{{\bf k}}-{\rm h.c}.  \nonumber
\end{eqnarray}
with the energy difference 
\begin{equation}
\alpha _{{\bf k,q}}=\varepsilon _{{\bf k+q}}-\varepsilon _{{\bf k}}+\omega _{%
{\bf q}}.  \label{EnDif}
\end{equation}
This generator yields several contributions to $dH(\ell )/d\ell =[\eta (\ell
),H(\ell )]$. The contribution to the change of $M_{{\bf k,q}}(\ell )$
results from $[\eta (\ell ),H_{0}]$

\begin{equation}
\frac{\partial M_{{\bf k,q}}(\ell )}{\partial \ell }=-\alpha _{{\bf k,q}%
}^{2}M_{{\bf k,q}}(\ell )  \label{Ex2FE1}
\end{equation}
with the solution

\begin{equation}
M_{{\bf k,q}}(\ell )=M_{{\bf q}}\exp (-\alpha _{{\bf k,q}}^{2}\ell ),
\label{FEMsol}
\end{equation}
where $M_{{\bf q}}$ is the initial electron-phonon coupling. For $\alpha _{%
{\bf k,q}}\neq 0$ the solution (\ref{FEMsol}) eliminates the electron-phonon
interaction from the Hamiltonian at $\ell \rightarrow \infty $. The
contribution to the effective electron-electron interaction is obtained from 
$[\eta (\ell ),H_{e-ph}(\ell )]$. For the electron pairs (with zero
momentum) we have the following {\it nonlinear} differential equation:

\begin{equation}
\frac{\partial V_{{\bf k,-k,q}}(\ell )}{\partial \ell }=-(\alpha _{{\bf k,q}%
}+\alpha _{-{\bf k-q,q}})M_{{\bf k,q}}(\ell )M_{-{\bf k-q,q}}(\ell )
\label{Ex2FE2}
\end{equation}
with the solution

\begin{equation}
V_{{\bf k,-k,q}}(\infty )=V_{{\bf k,-k,q}}(0)-\frac{M_{{\bf q}}^{2}\omega _{%
{\bf q}}}{\omega _{{\bf q}}^{2}+(\varepsilon _{{\bf k+q}}-\varepsilon _{{\bf %
k}})^{2}}.  \label{FEEEfinal}
\end{equation}
This result differs from (\ref{Vfinal}) (plus sign between the two squares
in the denominator). The difference is explained as: ``perturbation theory
for Hamiltonians is not uniquely defined'' \cite{wegner1}.

In the FE method a choice of $\eta (\ell )$ is motivated to be defined by $%
[H_{0},H_{e-ph}(\ell )]$ (\ref{EPHgen}). However, a rigorous criterion that
this choice (or, another one) is most optimal for a given problem seems has
not yet been formulated.

\section{Three-boson Hamiltonian}

In this example we consider the elimination of forbidden three-boson terms
from the Hamiltonian 
\begin{equation}
H(\ell )=\sum_{{\bf k}}\varepsilon _{{\bf k}}b_{{\bf k}}^{\dag }b_{{\bf k}%
}+H_{3}(\ell )+H_{4}(\ell ),  \label{ex3Ham}
\end{equation}
where

\begin{equation}
H_{3}(\ell )={\frac{1}{3}}\sum_{{\bf k},{\bf q}}\bigl\{\Psi _{{{\bf k},{\bf q%
}}}(\ell )b_{{\bf k}}^{\dag }b_{{\bf q}}^{\dag }b_{-{\bf k}-{\bf q}}^{\dag }+%
{\rm h.c.}\bigr\}  \label{ThreeBHam}
\end{equation}
and

\begin{eqnarray}
H_{4}(\ell ) &=&{\frac{1}{2}}\sum_{{{\bf k}_{1},{\bf k}_{2};{\bf k}_{3},{\bf %
k}_{4}}}\Phi _{{{\bf k}_{1},{\bf k}_{2};{\bf k}_{3},{\bf k}_{4}}}(\ell )b_{%
{\bf k}_{1}}^{\dag }b_{{\bf k}_{2}}^{\dag }b_{{\bf k}_{3}}b_{{\bf k}_{4}} 
\nonumber \\
&&\times \Delta ({{\bf k}_{1}+{\bf k}_{2}-{\bf k}_{3}-{\bf k}_{4}}).
\label{3BeffHam}
\end{eqnarray}
Here $\ell $ is a formal parameter and $\Delta ({\bf k)}$ is the Kronecker
delta function. For simplicity, we shall neglect the energy $\varepsilon _{%
{\bf k}}$ dependence on $\ell $.

Using OS CUT method, we write the parameter-independent operator in the form 
\cite{ShiSaf}: 
\begin{equation}
R=\sum_{{{\bf k},{\bf q}}}(R_{{{\bf k},{\bf q}}}b_{{\bf k}}^{\dag }b_{{\bf q}%
}^{\dag }b_{-{\bf k}-{\bf q}}^{\dag }-{\rm h.c.}).  \label{3Bgen}
\end{equation}

Calculating the commutators $[R,\sum_{{\bf k}}\varepsilon _{{\bf k}}b_{{\bf k%
}}^{\dag }b_{{\bf k}}]$ and $[R,H_{3}(\ell )]$, we obtain the following
linear differential equations:

\begin{equation}
\frac{d}{d\ell }\Psi _{{{\bf k},{\bf q}}}(\ell )=-3(\varepsilon _{{\bf k}%
}+\varepsilon _{{\bf q}}+\varepsilon _{{\bf -k-q}})R_{{{\bf k},{\bf q}}},
\label{3Bpsi}
\end{equation}
\begin{eqnarray}
\frac{d}{d\ell }\Phi _{{{\bf k}_{1},{\bf k}_{2};{\bf k}_{3},{\bf k}_{4}}%
}(\ell ) &=&-6R_{{{\bf k}_{1},{\bf k}_{2}}}\Psi _{{{\bf k}_{3},{\bf k}_{4}}%
}^{\ast }({\bf \ell })  \nonumber \\
&&-6R_{{{\bf k}_{3},{\bf k}_{4}}}^{\ast }\Psi _{{{\bf k}_{1},{\bf k}_{2}}%
}(\ell ).  \label{phi}
\end{eqnarray}
Solving these equations with the initial conditions $\Psi _{{{\bf k}_{1},%
{\bf k}_{2}}}(\ell =0)=\Psi _{{{\bf k}_{1},{\bf k}_{2}}}$, $\Phi _{{{\bf k}%
_{1},{\bf k}_{2};{\bf k}_{3},{\bf k}_{4}}}(\ell =0)=0$, and demanding $\Psi
_{{{\bf k}_{1},{\bf k}_{2}}}(\ell =1)=0$, we obtain 
\begin{equation}
R_{{{\bf k},{\bf q}}}=\frac{\Psi _{{{\bf k},{\bf q}}}}{3(\varepsilon _{{\bf k%
}}+\varepsilon _{{\bf q}}+\varepsilon _{{\bf -k-q}})}  \label{3Bgensol}
\end{equation}
and 
\begin{eqnarray}
\Phi _{{{\bf k}_{1},{\bf k}_{2};{\bf k}_{3},{\bf k}_{4}}}(1) &=&-\Psi _{{%
{\bf k}_{1},{\bf k}_{2}}}\Psi _{{{\bf k}_{3},{\bf k}_{4}}}^{\ast }\Big(\frac{%
1}{\varepsilon _{{\bf k}_{1}}+\varepsilon _{{\bf k}_{2}}+\varepsilon _{-{\bf %
k}_{1}-{\bf k}_{2}}}  \nonumber \\
&&+\frac{1}{\varepsilon _{{\bf k}_{3}}+\varepsilon _{{\bf k}%
_{4}}+\varepsilon _{-{\bf k}_{3}-{\bf k}_{4}}}\Big).  \label{3beffPhi}
\end{eqnarray}
It should be noted that this result for the effective four-boson interaction
coincides (with the additional condition $\Psi _{1}=\Psi _{1}^{\ast }$) with
the result of classical diagonalization of (\ref{ThreeBHam}) (when $b_{{\bf k%
}}^{\dag }$ and $b_{{\bf k}}$ are $c$-numbers) by canonical quasi-linear
transformation \cite{Krasit}.

Let us now consider the same problem in the framework of FE method. The
(parameter-dependent) generator in this case is:

\begin{equation}
\eta (\ell )=\frac{1}{3}\sum_{{\bf k},{\bf q}}\{\beta _{{\bf k,q}}\Psi _{{%
{\bf k},{\bf q}}}(\ell )b_{{\bf k}}^{\dag }b_{{\bf q}}^{\dag }b_{-{\bf k}-%
{\bf q}}^{\dag }-{\rm h.c.}\},  \label{3BFEgene}
\end{equation}
where

\begin{equation}
\beta _{{\bf k},{\bf q}}=\varepsilon _{{\bf k}}+\varepsilon _{{\bf q}%
}+\varepsilon _{-{\bf k}-{\bf q}}.  \label{beta}
\end{equation}

Corresponding differential equation for the three-boson amplitude has the
form:

\begin{equation}
\frac{d}{d\ell }\Psi _{{{\bf k},{\bf q}}}(\ell )=-\beta _{{\bf k},{\bf q}%
}^{2}\Psi _{{{\bf k},{\bf q}}}(\ell )  \label{3BFEPsieq}
\end{equation}
with the solution (compare with (\ref{FEMsol})):

\begin{equation}
\Psi _{{{\bf k},{\bf q}}}(\ell )=\Psi _{{{\bf k},{\bf q}}}\exp (-\beta _{%
{\bf k},{\bf q}}^{2}\ell ).  \label{3bFEsol}
\end{equation}

The differential equation for the four-boson effective amplitude in this
case is

\begin{equation}
\frac{d\Phi _{{{\bf k}_{1},{\bf k}_{2};{\bf k}_{3},{\bf k}_{4}}}(\ell )}{%
d\ell }=-2(\beta _{{\bf k}_{1}{\bf ,k}_{2}}+\beta _{{\bf k}_{3}{\bf ,k}%
_{4}})\Psi _{{{\bf k}_{1},{\bf k}_{2}}}(\ell )\Psi _{{{\bf k}_{3},{\bf k}_{4}%
}}^{\ast }(\ell ).  \label{3BFEeqPhi}
\end{equation}
The solution at $\ell \rightarrow \infty $ (when $\Psi _{{{\bf k},{\bf q}}%
}(\ell )$ vanishes) has the form

\begin{equation}
\Phi _{{{\bf k}_{1},{\bf k}_{2};{\bf k}_{3},{\bf k}_{4}}}(1)=-2\Psi _{{{\bf k%
}_{1},{\bf k}_{2}}}\Psi _{{{\bf k}_{3},{\bf k}_{4}}}^{\ast }\frac{\beta _{%
{\bf k}_{1}{\bf ,k}_{2}}+\beta _{{\bf k}_{3}{\bf ,k}_{4}}}{\beta _{{\bf k}%
_{1}{\bf ,k}_{2}}^{2}+\beta _{{\bf k}_{3}{\bf ,k}_{4}}^{2}}.
\label{3BFEPhifinal}
\end{equation}
We see that this formula differs from the OS CUT result (\ref{3beffPhi}) and
therefore (\ref{3BFEPhifinal}) does not correspond to the classical case 
\cite{Krasit}.

\section{Non-perturbative solution of quantum equations}

\subsection{Heisenberg picture}

The Heisenberg equation for an operator $A(t)$

\begin{equation}
i\hbar \frac{d}{dt}A(t)=[A(t),H(t)]  \label{Heisenberg}
\end{equation}
can be rewritten in the form:

\begin{equation}
\frac{d}{dt}A(t)=[iH(t)/\hbar ,A(t)].  \label{Heisenberg1}
\end{equation}
Formally this equation is similar to (\ref{equationUT}) (or, (\ref{FE}) for
time-dependent Hamiltonian): time $t$ plays a role of a formal parameter and 
$iH(t)/\hbar $ is an anti-Hermitian operator. As in the CUT method, we can
write the most general form of $A(t)$ as the expansion in terms of operator
combinations with unknown time-dependent coefficients. After substituting
these form into (\ref{Heisenberg1}) with a given $iH(t)/\hbar $, one can
obtain a set of differential equations by comparing coefficients in
analogous operator compositions in both sides. These equations will be
linear for time-independent $H$ and nonlinear for time-dependent $H(t)$.
Then we solve these equations with the initial conditions defined by $%
A(t)=A(0)$. Recently Mi\v{s}ta and Filip \cite{mista}\ have developed this
idea as a method of non-perturbative solutions of nonlinear Heisenberg
equations with an illustrative example of two coupled harmonic oscillators.

\subsection{Density matrix}

It is obvious that a quite analogous idea can be applied for the density
matrix. In this case we can write the density matrix equation as

\begin{equation}
\frac{d}{dt}\rho (t)=[-iH(t)/\hbar ,\rho (t)],  \label{density matrix}
\end{equation}
where $-iH(t)/\hbar $ plays a role of ``transformation generator''.

As an illustration consider a system of nuclear spins ${\bf I}_{j}$ with a
Hamiltonian $H=H_{0}+H_{int}$, where $H_{0}=\mu B\sum_{j}I_{j}^{z}$ is the
Zeeman energy and $H_{int}$ describes spin-spin interactions. As an initial
state assume the transverse magnetization. In the high temperature
approximation one can write $\rho (0)=1-\alpha \sum_{j}I_{j}^{x}$. Then the
density matrix can be represented as $\rho (t)=1-\sum_{n=1}^{\infty }\Delta
\rho _{n}(t)$, where $\Delta \rho _{n}(t)$ describes compositions of $n$
spins with time dependent coefficients, e.g., $\Delta \rho
_{1}(t)=\sum_{j}\left( X_{j}(t)I_{j}^{x}+Y_{j}(t)I_{j}^{y}\right) $. From $%
[H_{0},-\Delta \rho _{1}(t)]$ we obtain differential equations $%
dX_{j}(t)/dt=-\omega _{0}Y_{j}(t)$, $dY_{j}(t)/dt=\omega _{0}X_{j}(t)$ and
solutions $X_{j}(t)=\alpha \cos \omega _{0}t$, $Y_{j}(t)=\alpha \sin \omega
_{0}t$, where $\omega _{0}=\mu B/\hbar $ is the nuclear magnetic resonance
frequency \ Contributions to two spin motions can be found from $%
[H_{0},-\Delta \rho _{2}(t)]+[H_{int},-\Delta \rho _{1}(t)]$ and so on. This
example can demonstrate how the order stored in magnetization is transformed
to higher-order spin correlations.

\bigskip

The author wish to thank G. S. Uhrig and A. K. Khitrin\ for valuable
comments.


\begin{references}
\bibitem{wegner0}  F. Wegner, Ann. Physik {\bf 3}, 77 (1994).

\bibitem{keh}  S. K. Kehrein, A. Mielke, and P. Neu, Z. Phys. B {\bf 99},
269 (1996).

\bibitem{wegner1}  F. Wegner, Philosophical Magazine B {\bf 77}, 1249 (1998)

\bibitem{Stein0}  J. Stein, Eur. Phys. J. B {\bf 5}, 193 (1998).

\bibitem{moca}  C. P. Moca, I. Tifrea, and M. Crisan, J. of
Superconductivity {\bf 12}, 399 (1999).

\bibitem{Stein}  J. Stein, Europhys. Lett. {\bf 50}, 68 (2000).

\bibitem{Guba}  E. L. Gubankova, C.-R. Ji, and S. R. Cotanch, Phys. Rev. D 
{\bf 62}, 074001 (2000).

\bibitem{keh1}  W. Hofstetter and S. Kehrein, Phys. Rev. B {\bf 63},
140402(R) (2001).

\bibitem{knetter}  C. Knetter, K. P. Schmidt, M. Gr\"{u}ninger, and G. S.
Uhrig, Phys. Rev. Lett. 87, 167204 (2001).

\bibitem{Domanski}  T. Doma\'{n}ski, Eur. Phys. J. B {\bf 23}, 49 (2001).

\bibitem{raas}  C. Raas, A. B\"{u}hler, and G. S. Uhrig, Eur. Phys. J. B 
{\bf 21}, 369 (2001).

\bibitem{Jap}  Y. Ohira and K. Imafuku, quant-ph/ 0201005.

\bibitem{white}  S. R. White, cond-mat/0201346.

\bibitem{Ran}  T. Doma\'{n}ski and J. Ranninger, cond-mat/ 0202010.

\bibitem{Safonov0}  V. L. Safonov, Preprint KIAE - 3691/1, Moscow, August 8,
1982 (in Russian).

\bibitem{Safonov1}  V. L. Safonov, Phys. Lett. A {\bf 97}, 164 (1983).

\bibitem{SafNSW}  A. V. Andrienko, V. I. Ozhogin, V. L. Safonov, and A. Yu.
Yakubovskii, Zh. Eksp. Teor. Fiz. {\bf 84}, 1158 (1983) [Sov. Phys. JETP 
{\bf 57}, 673 (1983)].

\bibitem{SafFarz}  V. L. Safonov and R. M. Farzetdinova, J. Magn. Magn.
Mater. {\bf 98}{, }L235 (1991).

\bibitem{SafPara}  V. L. Safonov, Phys. Stat. Solidi (b) {\bf 174}{, 223}
(1992).

\bibitem{ShiSaf}  Q. Shi, V. L. Safonov, M. Mino, and H. Yamazaki, Phys.
Lett. A {\bf 238}, 258 (1998).

\bibitem{kittel}  C. Kittel, {\it Quantum Theory of Solids} (Wiley, New
York, 1963), p.152.

\bibitem{Krasit}  V. P. Krasitskii, Sov. Phys.-JETP {\bf 71,} 921 (1990).

\bibitem{mista}  L. Mi\v{s}ta, Jr. and R. Filip, J. Phys. A: Math. Gen. {\bf %
34, }5603 (2001).
\end{references}
\end{document}